\documentclass[prl, reprint, nofootinbib]{revtex4-1}
\usepackage{amsmath}
\usepackage{color}
\usepackage{graphicx}
\usepackage{tikz}

\newcommand{\abs}[1]{\left|#1\right|}
\newcommand{\Dx}{\mathcal Dx}
\newcommand{\Du}{\overline  D}
\newcommand{\ket}[1]{\left|#1\right>}
\newcommand{\cN}{\mathcal N}
\newcommand{\cD}{\mathcal D}

\renewcommand{\Im}{\text{Im}}

\newcommand{\nonlintime}{T_{\text{NL}}}
\newcommand{\sloshtime}{T_{\text{slosh}}}

\def\be{\begin{equation}}
\def\ee{\end{equation}}

\def\cO{\mathcal{O}}
\def\cD{\mathcal{D}}

\def\FV{\text{FV}}

\def\DV{\text{R}}
\def\R{\text{R}}
\def\omegaa{\omega_a}
\def\Ea{E_a}
\newcommand{\Veff}{V_{\text{eff}}}
\def\bigtau{\mathcal{T}}

\keywords{}
\begin{document}
\title{A direct approach to quantum tunneling}
\author{Anders Andreassen}
	\email{anders@physics.harvard.edu}
		\affiliation{Harvard University}
\author{David Farhi}
	\email{farhi@physics.harvard.edu}
		\affiliation{Harvard University}
\author{William Frost}
	\email{wfrost@physics.harvard.edu}
		\affiliation{Harvard University}
\author{Matthew D. Schwartz}
	\email{schwartz@physics.harvard.edu}
		\affiliation{Harvard University}
		
\date{\today}

\begin{abstract}
The decay rates of quasistable states in quantum field theories are usually calculated using instanton methods.
Standard derivations of these methods rely in a crucial way upon deformations and analytic continuations of the physical potential, and on the saddle point approximation.
While the resulting procedure can be checked against other semi-classical approaches in some one-dimensional cases, 
it is challenging to trace the role of the relevant physical scales, and any intuitive handle on the precision of the approximations involved are at best obscure.
In this paper, we use a physical definition of the tunneling probability to derive a formula for the decay rate in both quantum mechanics and quantum field theory directly from the Minkowski path integral,
without reference to unphysical deformations of the potential. There are numerous benefits to this approach,
from non-perturbative applications to precision calculations and aesthetic simplicity.
\end{abstract}

\maketitle

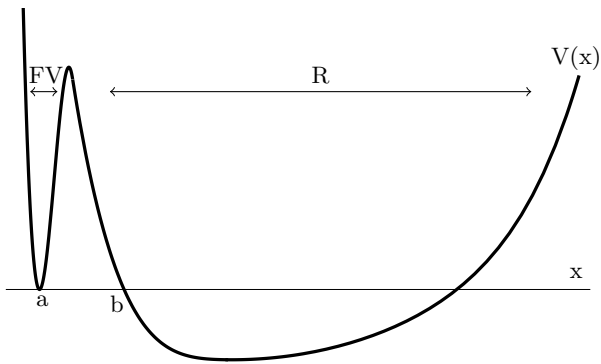
\begin{figure}[t]
\begin{center}
\begin{center}
\begin{tikzpicture}
\node at (0,0) {\includegraphics[width=0.9\columnwidth]{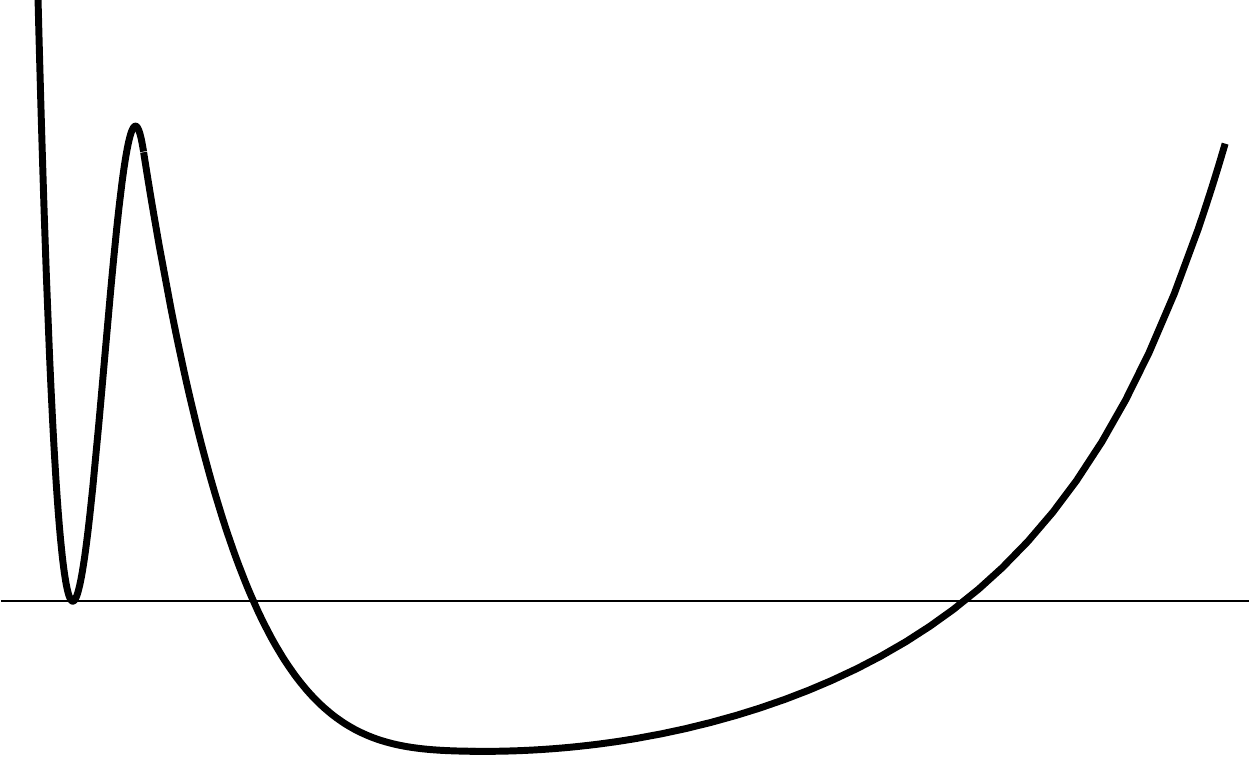}};
\draw [<->] (-3.56,1.3) -- (-3.2,1.3);
\node [above] at (-3.35,1.3) {\FV}; 
\draw [<->] (-2.5,1.3) -- (3.1,1.3);
\node [above] at (0.3,1.3) {\R};
\node [above] at (3.7,-1.3) {x};
\node at (3.7,1.75) {V(x)};
\node [below] at (-3.4,-1.3) {a};
\node [below left] at (-2.2,-1.3) {b};
\end{tikzpicture}
\end{center}
\end{center}
\caption{
An example 1D potential exhibiting quantum tunneling from the region $\FV$ to the region $\R$.}
\label{fig:potentialplot}
\end{figure}

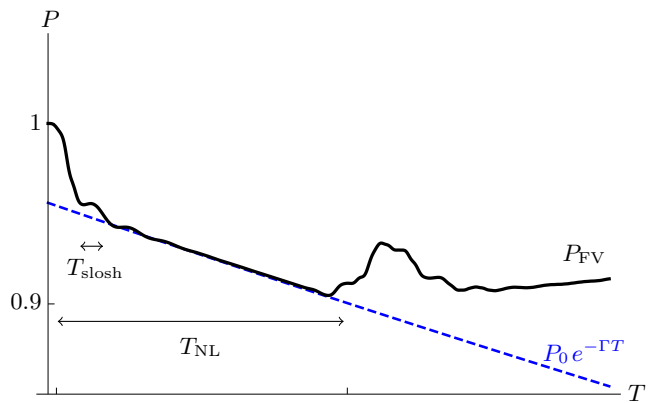
\begin{figure}[t]
\begin{center}
\begin{tikzpicture}
\node at (0,0) {\includegraphics[width=0.9\columnwidth]{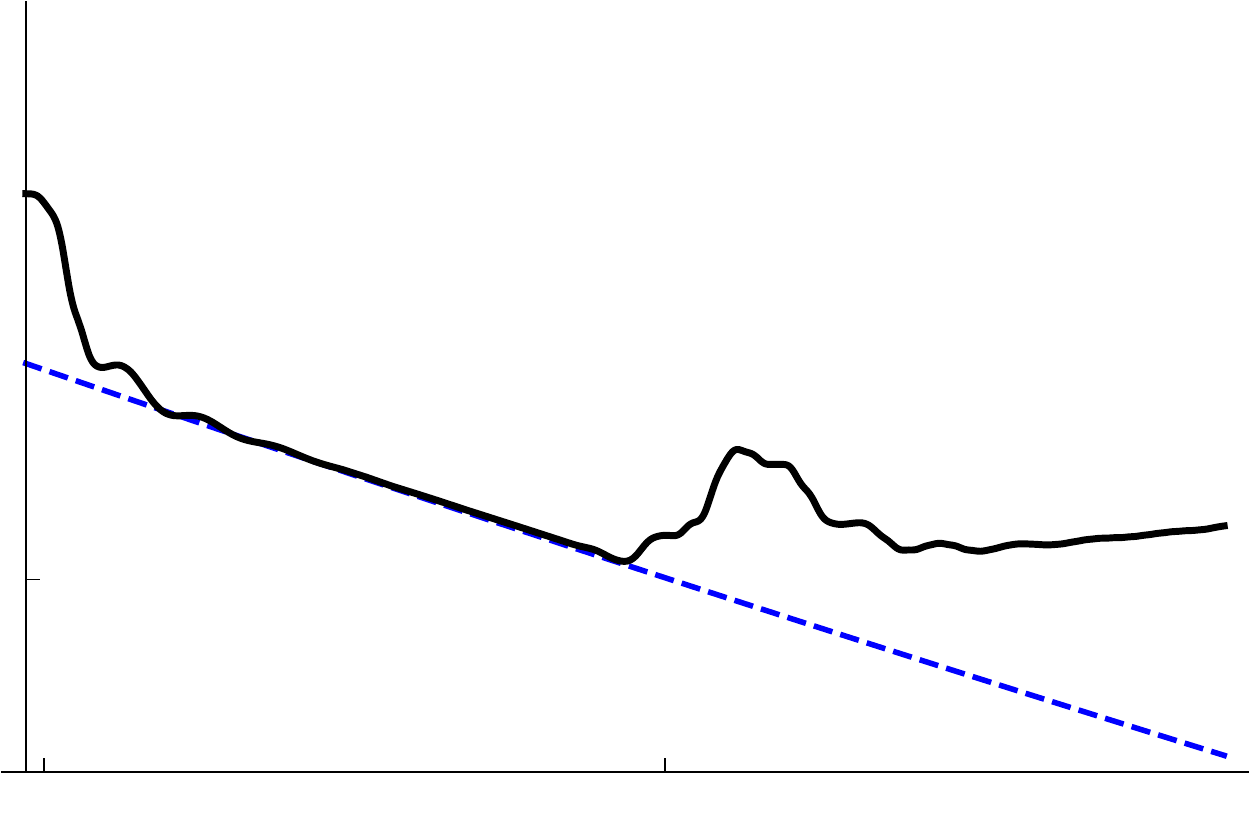}};
\node at (3.4,-0.4) {$P_{\FV}$}; 
\node at (3.4,-1.7) [blue] {$P_0\hspace{.3mm}e^{-\Gamma T}$};
\node at (4.1,-2.25) {$T$}; 
\draw [<->] (-3.3,-0.3) -- (-3.0,-0.3);
\node [below] at (-3.1,-0.35) {$\sloshtime$}; 
\draw [<->] (-3.6,-1.3) -- (0.2,-1.3);
\node [below] at (-1.7,-1.4) {$\nonlintime$}; 
\node [above] at (-3.7,2.5) {$P$}; 
\node [left] at (-3.7,1.35) {$1$}; 
\node [above left] at (-3.7,-1.25) {$0.9$}; 
\end{tikzpicture}
\end{center}
\caption{
The probability $P_{FV}(T)$ of funding a wavefunction in $\FV$ at time $t$
for the toy 1D potential in Fig.~\ref{fig:potentialplot}. This curve is computed by numerically integrating the Schr\"odinger equation beginning with a 
Gaussian wavepacket localized near $a$ at $t=0$. The false-vacuum probability falls exponentially, $P_{\FV}\sim e^{-\Gamma T}$, 
for times intermediate between the false-vacuum sloshing time $\sloshtime$ and the time when nonlinearities set in and the flux starts flowing back into the $\FV$ region.}
\label{fig:probabilityplot}
\end{figure}

Quantum tunneling is a hallmark of non-classical physics.
In 1D quantum mechanics the decay of quasistable states can be seen by solving the Schr\"odinger equation. An example is shown in 
Fig.~\ref{fig:potentialplot}. A wavefunction $\psi$, initially localized in the false-vacuum region (FV) near the point $a$, will evolve
in time to have support in the region labeled $\R$. When the lifetime is long, there is a well-defined
decay rate $\Gamma$. To see this, we begin with the probability of $\psi$ being found in region $\R$ after time $T$, given by:
\begin{equation}
\label{prob}
P_\R(T)\equiv\int_{\R} dx\abs{\psi(x,T)}^2
\end{equation}
This probability for the potential in Fig.~\ref{fig:potentialplot} is shown in Fig.~\ref{fig:probabilityplot} (computed numerically by solving the Schr\"odinger equation). Note that there are wiggles in the probability
 on the short timescale $\sloshtime\sim \omegaa^{-1}$, where $m\omegaa^2 = V''(a)$ characterizes the frequency of sloshing around the false vacuum. At intermediate
 times, the probability is exponential, $P_\FV(T) \sim e^{-\Gamma T}$, and we can extract $\Gamma$ from this regime.
 At late times non-linearities set in, both from the wavefunction bouncing off the far edge of $\DV$ and the initial wavefunction being depleted.

There is no practical way to generalize directly the above procedure, of numerically solving the Schr\"odinger equation, to compute decay
rates in quantum field theory.  There are alternative ways to compute the decay rate in 1D~\cite{razavy2013quantum}, such as the WKB approximation, or finding the imaginary part of resonance energies with Gamow-Siegert outgoing boundary conditions~\cite{Gamow:1928,siegert1939derivation}. However, one cannot easily justify the simplest generalizations of these to field theory either. The only approach that seems to generalize nicely is based on the path integral.

In the approach of Callan and Coleman~\cite{Callan:1977pt},  the decay rate is extracted from the  persistence amplitude in Euclidean time:
\begin{equation}
\label{Callan}
Z(T)=\left<a|e^{-HT}|a\right> = \sum_n e^{-E_n T} |\psi_n(a)|^2
\mathrel{\underset{T\to\infty}{\scalebox{2}[1]{$\sim$}}}
e^{-E_0 T}
\end{equation}
At large $T$, the sum is dominated by the ground state energy $E_0$, whose imaginary part gives half the decay rate. Writing the
matrix element in terms of a path integral, the claim is that
\begin{equation}
\label{ColemanPI}
\frac{\Gamma}{2} \sim  \Im \lim_{T\to\infty}\frac{1}{T} \ln \int_{x(-T/2)=a}^{x(T/2)=a} \cD x e^{-S[x]}
\end{equation}
A similar formula results from considering the partition function
which gives  a path integral with periodic rather than fixed boundary conditions~\cite{ZinnJustin:2002ru,Marino}. 

We write $\sim$ in Eq.~\eqref{ColemanPI} because one cannot simply compute the imaginary part:
the path integral produces $Z(T)$ of Eq.~\eqref{Callan}, which is manifestly real.
To get an imaginary part, most discussions typically take an example such as the quartic potential, $V(x)=x^2 -g x^4$.
For $g>0$, the spectrum for this potential is unbounded from below and $Z(T)$ is infinite. One can nevertheless get
a sensible answer for $\Im\,Z(T)$ with $g>0$ by analytic continuation from $g<0$, where the potential has a stable minimum.
The result for this case is that $Z(T)$ has a branch cut for $g>0$, with $\Im\,Z(T)$ given by $\pm \frac{1}{2}$ the discontinuity across the cut, which in turn is approximated by
the sum over saddle point expansions of $Z(T)$. 
These saddle points are finite-action instanton configurations called bounces.
Summing over these bounces in the dilute gas approximation, with the extra factor of $\frac{1}{2}$, gives a tunneling rate in quantum mechanics which agrees
with the formula from the WKB approximation to at least $\cO(\hbar^2)$~\cite{Jentschura:2011zza}.

The divergence of $Z(T)$ for the quartic case, following from the unboundedness of the potential, is not the general justification of the required potential deformation. In fact,
all physical potentials are bounded from below. For these, like $V(x)=x^2-gx^4+\lambda x^6$ or the one sketched in Figure~\ref{fig:potentialplot}, 
there is a ground state $E_0$ which is real. Then at large $T$, Eq.~\eqref{Callan} picks up the true ground state, associated with the minimum of region $\DV$. 
In the path integral, the dominant path is neither the bounce, nor the static solution $x(t)=a$ but rather a path which starts at $a$, quickly goes to the minimum
of region $\DV$, stays there for most of $T$, then returns to $a$. This path, which we call {\it the shot}, has very little to do with the decay, but is in fact the result of Eq.~\eqref{Callan}. 

To get an imaginary part in these cases, technically one must first deform the potential so that the false vacuum is absolutely stable,
then perform the saddle point approximation (adding by hand the factor of $\frac{1}{2}$ relating the analytic function to the sum over saddle points), then take the $T\to\infty$ 
limit\footnote{Note that the $T\to \infty$ limit must be done before analytically continuing back to the unstable case of interest. This can be seen if schematically $Z(T)=e^{-E_0(g)T}+e^{-\Ea(g)T}$, where $E_0$ is the energy near the true vacuum and $\Ea$ the energy near the false vacuum. Then it is clear that the $T\to\infty$ limit picks out $\min(E_0,\Ea)$, which changes non-analytically exactly when the false vacuum becomes unstable.}, and finally analytically continue the result back to the unstable case of interest. This procedure results  in an algorithm: the decay rate is given by the imaginary part of the instanton saddle point computed in the unstable case. 

Unfortunately, there seems to be no proof in the literature that the final prescription --- sum over bounces for the original potential with $T=\infty$ with a $\frac{1}{2}$ added by hand --- will always give the decay rate. What seems clear is that there
is a mathematically consistent way to define the imaginary part for the real quantity $Z(T)$ in the $T\to\infty$ limit. 
However, there is a surprising lack of commentary on the connection between this imaginary quantity and the physical decay rate.
Partly, this may be because much of the interest in the path integral formulation of tunneling is related to unraveling non-perturbative elements of quantum mechanics and quantum field theory.\footnote{
For example, although the quartic potential with $g>0$ is stable, the ground state energy as a function of $g$ is non-analytic, with zero radius of convergence. The finite-action instanton configurations for negative $g$ describe
the poles in the Borel transform of the positive $g$ series.} 
Partly, it may be because in quantum mechanics the final prescription can be proven equivalent to WKB at next-to-leading order (NLO),
 and there has been no 
reason to doubt the prescription in quantum field theory and no need for ultraprecise calculations. 
However, for questions about gauge-invariance of the decay rate~\cite{Metaxas:1995ab,Andreassen:2014gha,Andreassen:2014eha}, or for situations where the potential is itself generated by quantum effects, having a cleaner and more precise derivation may be helpful.
Moreover, since the derivation is so far removed from the physical problem, one cannot help but wonder if there might be a less byzantine connection between the path integral and the decay rate. 
It is the purpose of this paper to provide such a connection.  

From our initial discussion of the decay, as in Fig.~\ref{fig:probabilityplot}, we see that there are two characteristic times for physical potentials: a sloshing time $\sloshtime$ associated
with oscillations around the false vacuum and a non-linearity time, $\nonlintime$, associated with the structure of the potential around the true vacuum, on which timescales the picture of probability flowing linearly out of $\FV$ breaks down. 
The decay rate, which characterizes the linear fall during the intermediate time, can be extracted from Eq.~\eqref{prob} through the double limit
\begin{equation}
\label{defineG}
\Gamma \equiv 
-\hspace{-1em}\lim_{\substack{T/\nonlintime\to0\\T/\sloshtime\to\infty}} \frac{1}{P_{\FV}(T)}\frac{d}{dT}P_{\FV}(T)
\end{equation}
The limits in this definition are only mutually consistent
if $\nonlintime\gg \sloshtime$; i.e. the timescale for the system to leave the destination region and/or re-enter the false-vacuum region must be much longer than the timescales of sloshing around in the false vacuum. If this is not the case, then there is no use in characterizing any part of the probability function $P_\FV(T)$ with a ``decay rate."

Formula \eqref{defineG} easily generalizes to the case where there are multiple directions for decay. In that situation we might be interested 
in the total decay rate $\Gamma$ as well as 
the decay rate  $\Gamma_R$ to a particular region $R$ of configuration space, which is not necessarily the entire compliment of the false vacuum region.
Then we should look not at the fractional linear fall-off of  $P_{\FV}$, but rather the fractional linear rise of $P_\R$, the probability to find the system in $\R$.
Then
\begin{equation}
\label{defineGR}
\Gamma_\R \equiv 
\hspace{-1em}
\lim_{\substack{T/\nonlintime^\R \to0\\T/\sloshtime\to\infty}} \frac{1}{P_{\FV}(T)}\frac{d}{dT}P_{\R}(T)
\end{equation}
where now $\nonlintime^\R$ is the timescale for the system to leave $\R$ (back to $\FV$ or to any other wells that may be present). The above definition can be recast in terms of the conserved quantum mechanical probability flux if desired.

For potentials of interest, where $\nonlintime^\R \gg \sloshtime$, the exact energy eigenstates that have substantial support in the $\FV$ region are in bands around resonance energies $E_i$ with widths $\Gamma_i$.
The higher energy bands will decay exponentially faster than the lower energy ones, so we will focus on the lowest energy band. 
Any wavefunction with support on this band will have the same decay rate.
Thus for simplicity we can begin with a position-eigenstate localized at exactly the minimum of the false vacuum well:
$\psi(x,0)=\delta(x-a)$. For $t>0$, the wavefunction is simply the propagator and so
\begin{align}
\label{PfromPI}
P_\R(T) 
& = \int_\R dx_f\abs{\cN D(a, 0; x_f, T)}^2 
\\
& = \int_\R dx_f\abs{\cN \int_{x(0)=a}^{x(T)=x_f} \Dx e^{iS[x]}}^2
\end{align}
For later convenience, we have written the propagator as $\cN D$ rather than simply $D$, where $\cN$ is the path integral normalization defined by the second line.

To proceed, we will use the identity that, for any point $b$ between $a$ and $x_f$, 
we can label the {\it first} time the path hits $b$ by $t_0$ and write
\begin{equation}
\label{splitpath}
D(a, 0; x_f, T)  \equiv \int_0^T dt_0 \Du(a, 0; b, t_0)D(b, t_0; x_f, T)
\end{equation}
where 
\begin{equation}
\label{defineDu}
\Du(a, 0; b, t_0)  = \int_{x(0)=a}\Dx e^{iS[x]}\delta(t_b[x]-t_0)
\end{equation}
with $t_b[x]$ the functional that returns the first time a path $x(t)$ hits the point $b$ and $\delta(t_b[x]-t_0)=0$ if $x(t)$ never hits $b$. Equivalently
\begin{equation}
\Du(a, 0; b, t_0) 
\label{defineDuBCs}
= \int_{\substack{x(0)=a\\x(t_0)=b\\x(t)<b}}\Dx e^{iS[x]}\abs{\dot x(t_0)}
\vspace{-1mm}
\end{equation}

Inserting Eq.~\eqref{splitpath} twice into Eq.~\eqref{PfromPI} gives
\begin{multline}
\label{fourDs}
\hspace{-6mm}
P_\R(T) = \cN \cN^*\int dt_0 \int d t_0' \int_\R d x_f   
\Du(a,0;b,t_0) \Du^{\,*}\!(a,0;b,t_0')\\
\times  D(b, t_0; x_f, T)D^*(b, t_0'; x_f, T)
\end{multline}

Now, recall the condition $T\ll\nonlintime^\R$ in Eq.~\eqref{defineGR}.
In the limit $T/\nonlintime^\R \to 0$,  no flux enters back into the $\FV$ region. Since $x_f$ only appears in the propagators 
to and from $b$, this condition allows us to replace the integral of $x_f$ over $\R$ with an integral of $x_f$ over all of configuration space;
the added probability from the propagator to points outside of $\R$ is negligible.
With the restriction to $\R$ removed, we can use the completeness of $\ket{x_f}$ to combine the two propagators in the second
line of Eq.~\eqref{fourDs} into a single $D(b, t_0; b, t_0')$, which can then be recombined with the remaining two $\Du$ factors using Eq.~\eqref{splitpath} in reverse. This leads to
\begin{equation}
\label{PRT}
P_\R(T) =\cN\cN^*\int_0^T dt_0D(a, 0; b, t_0)\Du^*(a, 0; b, t_0)+\text{c.c} 
\end{equation}

Finally, we find $\Gamma_\R$ by differentiating with respect to $T$, dividing by $P_{\FV}$, and taking $T\to \infty$.
 \begin{equation}
\label{gammaRMinkowski}
\boxed{
\Gamma_\R = 
\lim_{T\to\infty}
\frac
{D(a, 0; b, T)\Du^*(a, 0; b, T)+\text{c.c}}
{\int dx \abs{D(a, 0; x, T)}^2}
}
\end{equation}
Taking $T\to \infty$ enforces the other
 limit $T\gg \sloshtime$ in Eq.~\eqref{defineGR}, which is not in conflict with the $T\ll \nonlintime^\R$ limit since we have already enforced
 that limit to get to Eq.~\eqref{PRT}. This is an exact non-perturbative definition of the decay rate indepedent of the saddle-point approximation.

To make sense of Eq.~\eqref{gammaRMinkowski}, let us check that it agrees with existing results at NLO. First, we Wick
rotate to imaginary time. This provides a further simplification since the two remaining propagators can then be combined:
\begin{equation}
\label{gammaR}
\Gamma_\R = 
2\Im\lim_{T\to\infty}\left[ \frac
{\int
\Dx e^{-S_E[x]}\delta(\tau_b[x])}
{\int \Dx
  e^{-S_E[x]}}\right]_{\bigtau \to  i T}
\end{equation}
Where both path integrals have boundary condition $x(\pm\bigtau)=a$. The functional $\tau_b[x]$  is identical to $t_b[x]$, but we have renamed it as a reminder that $x$ is now a function of Euclidean time $\tau$. 
The ``$\Im$'' arises because the $\delta$ or $\dot x$ in $\Du$ in Eq.~\eqref{gammaRMinkowski}, with dimensions of time, acquires an $i$, which causes a relative sign between the two complex conjugate terms.

To proceed with the NLO expansion, we locate the extrema of the Euclidean action. Because of the $\delta$-function, all paths must hit $b$ at $\tau=0$. 
Thus the $x(\tau)=a$ extremum is removed but the bounce $\bar x (\tau)$ remains.\footnote{
Although the shot is removed as well, there is a modified shot $x_s(\tau)$ which is identical to the bounce until it hits $b$ and then shoots into $\R$ eventually returning to $a$. 
This modified shot has Euclidean action $S_E[x_s] \approx - \bigtau |E_0| + S_1$ with $S_1$ coming from when the shot is moving fast. 
Although $S_E[x_s] \ll S_E[\bar x]$ at large real $\bigtau$, after $\bigtau \to i T$, the contribution of modified shot is exponentially suppressed since $S_1 > S_E[\bar x]$. Thus
the bounce is the dominant contribution to the rate.
}
 A minor pleasing feature of this derivation is that approximate solutions, such as time-shifted bounces or multiple bounces, play no role.

To  integrate over the $\delta$-function carefully, we use the collective coordinate $\tau_0$.
Fluctuations around the bounce are parameterized by $\tau_0$ and $\{\zeta_i\}_{i>0}$, as $x(\tau,\tau_0,\zeta_i) = \bar x (\tau - \tau_0) + \sum_{i> 0} \zeta_i x_i(\tau - \tau_0)$
so that $\tau_b[x]=\tau_0 + \tau_b[x(\tau,0,\zeta_i)]$.
At NLO, the Jacobian generated by changing to $\{\tau_0,\zeta_i\}$ is $J=\sqrt{S_E(\bar x)/m}$~\cite{Callan:1977pt}.

The action is independent of $\tau_0$; hence the integral over $\tau_0$ is equivalent to multiplication by $J$ if the path $x(\tau,0,\zeta_i)$ hits $b$ and 0 otherwise.
This leaves the remaining Gaussian fluctuations with the extra boundary condition that all paths must hit $b$. 
Around $\bar x(t)$, which hits $b$ at its maximum, this restriction adds $\Theta[\zeta_ix_i(0)]$ to our path integral. At NLO the action is symmetric in the $\zeta$ fluctuations, so this restriction to half of the domain can be removed if a factor of $\frac{1}{2}$ is added.
This leaves a product of Gaussian integrals over the $\zeta$, whose result is

\begin{equation}
\label{GammaNLO}
\Gamma_\R = e^{-S[\bar x]} \sqrt{\frac{S_E[\bar x]}{2\pi}}\sqrt{\frac{\det\left[-\partial^2_\tau+V''(a)\right]}{-\det'\left[-\partial^2_\tau+V''(\bar x(\tau))\right]}}
\end{equation}
Here $\det'$ is the usual Fredholm determinant with the zero mode removed, as in~\cite{Callan:1977pt}. The minus sign comes from the negative-eigenvalue fluctuation which is still present
and contributes to the determinant without ado.
This formula is identical to the NLO formula following from Eq.~\eqref{ColemanPI} and agrees with the WKB approximation.

The above derivation generalizes naturally to multiple dimensional quantum mechanics and to field theory. The only difference is that one needs to integrate $b$ over a surface $\Sigma$ that bounds the destination region $\R$ (where $\Sigma$ and $\R$ are now regions in field configuration space).
In field theory, Eq.~\eqref{gammaR} becomes
\begin{equation}
\label{gammaRMinkowskiQFT}
\frac{\Gamma_\R}{V} = \frac{1}{V}
2\Im\frac
{\int \cD\phi e^{-S_E[\phi]}\delta(\tau_\Sigma[\phi])}
{\int \cD\phi e^{-S_E[\phi]}}
\end{equation}
Here the path integrals have boundary conditions $\phi(\vec x, \bigtau=\pm\infty)=\phi_a$, where $\phi_a$ is the false vacuum, and the same analytic continuation and limit as in Eq.~\eqref{gammaR} are understood. 
The functional $\tau_\Sigma[\phi(\vec x,\tau)]$  
returns the first time $\tau$ at which the configuration $\phi(\vec x, \tau)$ hits the surface $\Sigma$.

Often the natural choice for $\Sigma$ is the {\it turning point surface}: the set of configurations $\phi(\vec x)$ satisfying $U[\phi(\vec{x})]=U[\phi_a]$ where the classical energy functional is
\begin{equation}
\label{U0}
 U[\phi(\vec x)] = \int d^3x\left[ \frac{1}{2}(\vec\nabla\phi)^2+V(\phi)\right]
\end{equation}
Solving for $\Sigma$ will usually indicate one or more regions $\R$ disconnected from FV where Eq.~\eqref{gammaRMinkowskiQFT} can be used.

An important case where $\Sigma$ cannot be computed directly from $V$ is when tunneling is due to radiative corrections~\cite{Coleman:1973jx}. For example, in the Standard Model, the leading order Higgs potential $V(h) \sim \lambda h^4$ with $\lambda>0$ at the weak scale.
The $U=0$ surface is then simply $h=0$ and there is no tunneling. But at high energy, $\lambda <0$ and there is tunneling. This tunneling can be seen if the effective potential $\Veff$ is
used instead of $V$ in Eq.~\eqref{U0}. However, using $\Veff$ to compute tunneling rates is incorrect: one cannot compute the path integral using $\Veff$ without double counting.
The formulation we described provides an alternative: one can use
Eq.~\eqref{gammaRMinkowskiQFT} with the classical potential $V$ to compute the tunneling rate. 

To conclude, let us summarize some distinctions between the direct approach described here and the conventional potential-deformation method. 
There are many small technical differences, such as the irrelevance of approximate bounce solutions, multiple bounces, and the dilute gas
approximation, but also some larger conceptual differences.

First, in our derivation, the role of the two relevant time scales for the decay to be well-defined, $\nonlintime$ and $\sloshtime$, is undeniable. Taking $T \ll \nonlintime$ prevents tunneling back into the false-vacuum region.
The limit lets us approximate the integral over $\R$ as an integral over all space, and is an exponentially-small unitarity-violating approximation morally equivalent to the Gamow-Siegert outgoing boundary conditions.
 In the potential-deformation method, $\nonlintime$ is never invoked. It does, however, also play a critical role there as well: because $\nonlintime$ is finite, one must analytically continue the potential so that the $\FV$ is absolutely stable, then take $T \to \infty$, then analytically continue back. This is how the shot contribution is avoided, and why one {\it must} analytically continue the potential to get the decay
rate from the partition function.

A second important difference is that in our approach the only analytic continuation used is the usual Wick rotation to imaginary
time.
When the potential is analytically continued,
the sum over instantons gives the discontinuity across the cut, which differs from the naive imaginary part by a factor of $\frac{1}{2}$. In our case, the
factor of $\frac{1}{2}$  
arises because only half of the fluctuations around the bounce reach $b$.

Not having to analytic continue the potential implies that one could also, in principle at least, evaluate the decay rate numerically in a quantum field theory. 
For example, one could us  Eq.~\eqref{gammaRMinkowski}, in principle, to compute the rate for $\alpha$ decay from first principles in QCD on the lattice.

Finally, it is worth recalling that the generalization of the Callan-Coleman approach to the multi-dimensional case is often done with a leap of faith. Partly, this is because it is meant to agree with WKB and the generalization of
WKB to decays in more than one dimension is complicated~\cite{Banks:1973ps}; partly it is because studying the analytic continuation of a energy functional of fields with multiple stable regions is grossly complicated.
In contrast, the direct derivation described here is fundamentally path-integral based so multiple dimensional case and field theory versions are essentially identical.

We thank D. Harlow, M. Mari\~{n}o, A. Strumia, M. Unsal and E. Weinberg for helpful discussions. This research was supported in part by the U.S. Department of Energy, under grant DE-SC0013607.

\bibliography{all-orders-decay}

\begin{thebibliography}{12}%
\makeatletter
\providecommand \@ifxundefined [1]{%
 \@ifx{#1\undefined}
}%
\providecommand \@ifnum [1]{%
 \ifnum #1\expandafter \@firstoftwo
 \else \expandafter \@secondoftwo
 \fi
}%
\providecommand \@ifx [1]{%
 \ifx #1\expandafter \@firstoftwo
 \else \expandafter \@secondoftwo
 \fi
}%
\providecommand \natexlab [1]{#1}%
\providecommand \enquote  [1]{``#1''}%
\providecommand \bibnamefont  [1]{#1}%
\providecommand \bibfnamefont [1]{#1}%
\providecommand \citenamefont [1]{#1}%
\providecommand \href@noop [0]{\@secondoftwo}%
\providecommand \href [0]{\begingroup \@sanitize@url \@href}%
\providecommand \@href[1]{\@@startlink{#1}\@@href}%
\providecommand \@@href[1]{\endgroup#1\@@endlink}%
\providecommand \@sanitize@url [0]{\catcode `\\12\catcode `\$12\catcode
  `\&12\catcode `\#12\catcode `\^12\catcode `\_12\catcode `\%12\relax}%
\providecommand \@@startlink[1]{}%
\providecommand \@@endlink[0]{}%
\providecommand \url  [0]{\begingroup\@sanitize@url \@url }%
\providecommand \@url [1]{\endgroup\@href {#1}{\urlprefix }}%
\providecommand \urlprefix  [0]{URL }%
\providecommand \Eprint [0]{\href }%
\providecommand \doibase [0]{http://dx.doi.org/}%
\providecommand \selectlanguage [0]{\@gobble}%
\providecommand \bibinfo  [0]{\@secondoftwo}%
\providecommand \bibfield  [0]{\@secondoftwo}%
\providecommand \translation [1]{[#1]}%
\providecommand \BibitemOpen [0]{}%
\providecommand \bibitemStop [0]{}%
\providecommand \bibitemNoStop [0]{.\EOS\space}%
\providecommand \EOS [0]{\spacefactor3000\relax}%
\providecommand \BibitemShut  [1]{\csname bibitem#1\endcsname}%
\let\auto@bib@innerbib\@empty
\bibitem [{\citenamefont {Razavy}(2013)}]{razavy2013quantum}%
  \BibitemOpen
  \bibfield  {author} {\bibinfo {author} {\bibfnamefont {M.}~\bibnamefont
  {Razavy}},\ }\href {https://books.google.com.au/books?id=AcJ0nQEACAAJ} {\emph
  {\bibinfo {title} {Quantum Theory of Tunneling}}}\ (\bibinfo  {publisher}
  {World Scientific Publishing Company Pte Limited},\ \bibinfo {year}
  {2013})\BibitemShut {NoStop}%
\bibitem [{\citenamefont {Gamow}(1928)}]{Gamow:1928}%
  \BibitemOpen
  \bibfield  {author} {\bibinfo {author} {\bibfnamefont {G.}~\bibnamefont
  {Gamow}},\ }\href@noop {} {\bibfield  {journal} {\bibinfo  {journal} {Z.
  Phys.}\ }\textbf {\bibinfo {volume} {51}},\ \bibinfo {pages} {204} (\bibinfo
  {year} {1928})}\BibitemShut {NoStop}%
\bibitem [{\citenamefont {Siegert}(1939)}]{siegert1939derivation}%
  \BibitemOpen
  \bibfield  {author} {\bibinfo {author} {\bibfnamefont {A.~J.}\ \bibnamefont
  {Siegert}},\ }\href@noop {} {\bibfield  {journal} {\bibinfo  {journal}
  {Physical Review}\ }\textbf {\bibinfo {volume} {56}},\ \bibinfo {pages} {750}
  (\bibinfo {year} {1939})}\BibitemShut {NoStop}%
\bibitem [{\citenamefont {Callan}\ and\ \citenamefont
  {Coleman}(1977)}]{Callan:1977pt}%
  \BibitemOpen
  \bibfield  {author} {\bibinfo {author} {\bibfnamefont {C.~G.}\ \bibnamefont
  {Callan}, \bibfnamefont {Jr.}}\ and\ \bibinfo {author} {\bibfnamefont
  {S.~R.}\ \bibnamefont {Coleman}},\ }\href {\doibase 10.1103/PhysRevD.16.1762}
  {\bibfield  {journal} {\bibinfo  {journal} {Phys. Rev.}\ }\textbf {\bibinfo
  {volume} {D16}},\ \bibinfo {pages} {1762} (\bibinfo {year}
  {1977})}\BibitemShut {NoStop}%
\bibitem [{\citenamefont {Zinn-Justin}(2002)}]{ZinnJustin:2002ru}%
  \BibitemOpen
  \bibfield  {author} {\bibinfo {author} {\bibfnamefont {J.}~\bibnamefont
  {Zinn-Justin}},\ }\href@noop {} {\bibfield  {journal} {\bibinfo  {journal}
  {Int. Ser. Monogr. Phys.}\ }\textbf {\bibinfo {volume} {113}},\ \bibinfo
  {pages} {1} (\bibinfo {year} {2002})}\BibitemShut {NoStop}%
\bibitem [{\citenamefont {Mari\~{n}o}(2015)}]{Marino}%
  \BibitemOpen
  \bibfield  {author} {\bibinfo {author} {\bibfnamefont {M.}~\bibnamefont
  {Mari\~{n}o}},\ }\href
  {http://www.cambridge.org/mw/academic/subjects/physics/theoretical-physics-and-mathematical-physics/instantons-and-large-n-introduction-non-perturbative-methods-quantum-field-theory?format=HB}
  {\emph {\bibinfo {title} {{Instantons and Large N: An Introduction to
  Non-Perturbative Methods in Quantum Field Theory}}}}\ (\bibinfo  {publisher}
  {Cambridge University Press},\ \bibinfo {year} {2015})\BibitemShut {NoStop}%
\bibitem [{\citenamefont {Jentschura}\ and\ \citenamefont
  {Zinn-Justin}(2011)}]{Jentschura:2011zza}%
  \BibitemOpen
  \bibfield  {author} {\bibinfo {author} {\bibfnamefont {U.~D.}\ \bibnamefont
  {Jentschura}}\ and\ \bibinfo {author} {\bibfnamefont {J.}~\bibnamefont
  {Zinn-Justin}},\ }\href {\doibase 10.1016/j.aop.2011.04.002} {\bibfield
  {journal} {\bibinfo  {journal} {Annals Phys.}\ }\textbf {\bibinfo {volume}
  {326}},\ \bibinfo {pages} {2186} (\bibinfo {year} {2011})}\BibitemShut
  {NoStop}%
\bibitem [{\citenamefont {Metaxas}\ and\ \citenamefont
  {Weinberg}(1996)}]{Metaxas:1995ab}%
  \BibitemOpen
  \bibfield  {author} {\bibinfo {author} {\bibfnamefont {D.}~\bibnamefont
  {Metaxas}}\ and\ \bibinfo {author} {\bibfnamefont {E.~J.}\ \bibnamefont
  {Weinberg}},\ }\href {\doibase 10.1103/PhysRevD.53.836} {\bibfield  {journal}
  {\bibinfo  {journal} {Phys.Rev.}\ }\textbf {\bibinfo {volume} {D53}},\
  \bibinfo {pages} {836} (\bibinfo {year} {1996})},\ \Eprint
  {http://arxiv.org/abs/hep-ph/9507381} {arXiv:hep-ph/9507381 [hep-ph]}
  \BibitemShut {NoStop}%
\bibitem [{\citenamefont {Andreassen}\ \emph {et~al.}(2014)\citenamefont
  {Andreassen}, \citenamefont {Frost},\ and\ \citenamefont
  {Schwartz}}]{Andreassen:2014gha}%
  \BibitemOpen
  \bibfield  {author} {\bibinfo {author} {\bibfnamefont {A.}~\bibnamefont
  {Andreassen}}, \bibinfo {author} {\bibfnamefont {W.}~\bibnamefont {Frost}}, \
  and\ \bibinfo {author} {\bibfnamefont {M.~D.}\ \bibnamefont {Schwartz}},\
  }\href {\doibase 10.1103/PhysRevLett.113.241801} {\bibfield  {journal}
  {\bibinfo  {journal} {Phys. Rev. Lett.}\ }\textbf {\bibinfo {volume} {113}},\
  \bibinfo {pages} {241801} (\bibinfo {year} {2014})},\ \Eprint
  {http://arxiv.org/abs/1408.0292} {arXiv:1408.0292 [hep-ph]} \BibitemShut
  {NoStop}%
\bibitem [{\citenamefont {Andreassen}\ \emph {et~al.}(2015)\citenamefont
  {Andreassen}, \citenamefont {Frost},\ and\ \citenamefont
  {Schwartz}}]{Andreassen:2014eha}%
  \BibitemOpen
  \bibfield  {author} {\bibinfo {author} {\bibfnamefont {A.}~\bibnamefont
  {Andreassen}}, \bibinfo {author} {\bibfnamefont {W.}~\bibnamefont {Frost}}, \
  and\ \bibinfo {author} {\bibfnamefont {M.~D.}\ \bibnamefont {Schwartz}},\
  }\href {\doibase 10.1103/PhysRevD.91.016009} {\bibfield  {journal} {\bibinfo
  {journal} {Phys. Rev.}\ }\textbf {\bibinfo {volume} {D91}},\ \bibinfo {pages}
  {016009} (\bibinfo {year} {2015})},\ \Eprint {http://arxiv.org/abs/1408.0287}
  {arXiv:1408.0287 [hep-ph]} \BibitemShut {NoStop}%
\bibitem [{\citenamefont {Coleman}\ and\ \citenamefont
  {Weinberg}(1973)}]{Coleman:1973jx}%
  \BibitemOpen
  \bibfield  {author} {\bibinfo {author} {\bibfnamefont {S.~R.}\ \bibnamefont
  {Coleman}}\ and\ \bibinfo {author} {\bibfnamefont {E.~J.}\ \bibnamefont
  {Weinberg}},\ }\href {\doibase 10.1103/PhysRevD.7.1888} {\bibfield  {journal}
  {\bibinfo  {journal} {Phys. Rev.}\ }\textbf {\bibinfo {volume} {D7}},\
  \bibinfo {pages} {1888} (\bibinfo {year} {1973})}\BibitemShut {NoStop}%
\bibitem [{\citenamefont {Banks}\ \emph {et~al.}(1973)\citenamefont {Banks},
  \citenamefont {Bender},\ and\ \citenamefont {Wu}}]{Banks:1973ps}%
  \BibitemOpen
  \bibfield  {author} {\bibinfo {author} {\bibfnamefont {T.}~\bibnamefont
  {Banks}}, \bibinfo {author} {\bibfnamefont {C.~M.}\ \bibnamefont {Bender}}, \
  and\ \bibinfo {author} {\bibfnamefont {T.~T.}\ \bibnamefont {Wu}},\ }\href
  {\doibase 10.1103/PhysRevD.8.3346} {\bibfield  {journal} {\bibinfo  {journal}
  {Phys. Rev.}\ }\textbf {\bibinfo {volume} {D8}},\ \bibinfo {pages} {3346}
  (\bibinfo {year} {1973})}\BibitemShut {NoStop}%
\end{thebibliography}%
\end{document}